\documentclass[twocolumn]{jpsj2} 

\def\lsco{La$_{2-x}$Sr$_x$CuO$_4$}
\def\lbco{La$_{2-x}$Ba$_x$CuO$_4$}
\def\lnsco{La$_{1.6-x}$Nd$_{0.4}$Sr$_x$CuO$_4$}
\def\ybco{YBa$_2$Cu$_3$O$_{6+x}$}

\title{Magnetic neutron scattering in hole doped cuprate superconductors}

\author{R.J. \textsc{Birgeneau}$^{1}$, C.\textsc{Stock}$^{2}$, J.M. \textsc{Tranquada}$^{3}$ , K.\textsc{Yamada}$^{4}$ }

\inst{$^{1}$  Department of Physics, University of California at Berkeley, Berkeley, California, 94729, USA  \\
$^{2}$ Department of Physics, Johns Hopkins University, Baltimore, Maryland, 21218, USA \\
$^{3}$ Physics Department, Brookhaven National Laboratory,  Upton, New York 11973, USA\\
$^{4}$Institute of Materials Research, Tohoku University, Sendai 980-8577, Japan}

\abst{A review is presented of the static and dynamic magnetic properties of hole-doped cuprate superconductors measured with neutron scattering.  A wide variety of experiments are described with emphasis on the monolayer La$_{2-x}$(Sr,Ba)$_{x}$CuO$_{4}$ and bilayer YBa$_{2}$Cu$_{3}$O$_{6+x}$ cuprates.  At zero hole doping, both classes of materials are antiferromagnetic insulators with large superexchange constants of J $>$ 100 meV.  For increasing hole doping, the cuprates become superconducting at a critical hole concentration of $x_{c}$=0.055.  The development of new instrumentation at neutron beam sources coupled with the improvement in materials has lead to a better understanding of these materials and the underlying spin dynamics over a broad range of hole dopings.  We will describe how the spin dispersion changes across the insulating to superconducting boundary as well as the static magnetic properties which are directly coupled with the superconductivity.  Experiments directly probing the competing magnetic and superconducting order parameters involving magnetic fields, impurity doping, and structural order will be examined.  Correlations between superconductivity and magnetism will also be discussed.}

\kword{superconductivitiy, neutron scattering}
\begin{document}
\maketitle

\section{Introduction}

	Magnetism plays an important role in the phase diagram of the high-temperature superconducting cuprates.  A complete understanding of high-temperature superconductivity must involve a thorough characterization and knowledge of the spin response across the entire phase diagram.~\cite{Kastner98:70,Kivelson03:75}  The magnetic properties throughout the phase diagram have proven to be immensely complex and sensitive to effects of disorder.  Currently, there is no complete theory of the spin spectrum able to describe the measured response throughout the entire phase diagram it is also not understood how the magnetism relates to the superconducting properties.  

	The cuprate compounds at zero charge doping are antiferromagnetic insulators which consist of two-dimensional planes of magnetic CuO$_{2}$ plaquettes coupled through a strong superexchange interaction.~\cite{Lee06:78} With one hole per Copper atom, one would expect that the undoped cuprate would be metallic and many calculations have indeed made this prediction.  However, all undoped cuprates are antiferromagnetic insulators with large N\'{e}el temperatures, illustrating the importance that electron-electron interactions play in the phase diagram.  High-temperature superconductivity is introduced in these materials through the doping of holes into the CuO$_{2}$ planes which causes a dramatic suppression of the long-ranged antiferromagnetism and at a critical hole doping results in a superconducting state. For very large hole dopings, superconductivity is suppressed in favor of a metallic ground state well described by Fermi-liquid theory. 

\begin{figure}[t]
\begin{center}
\includegraphics[width=7.5cm] {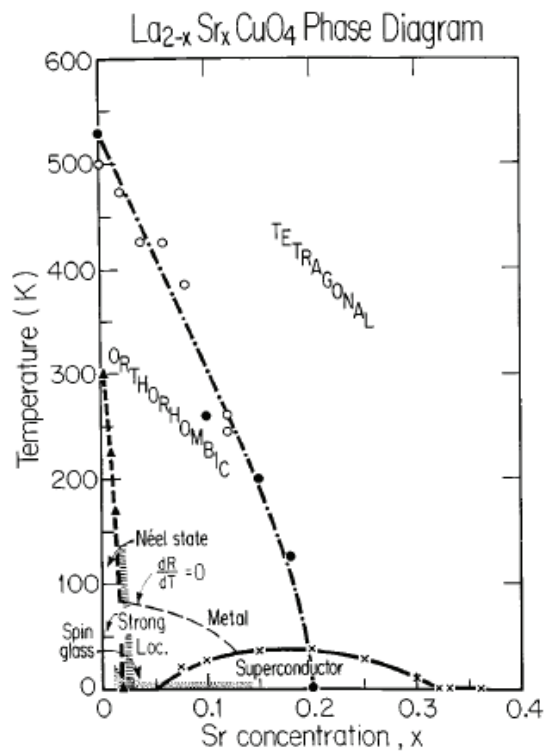}
\end{center}
\caption{\label{phase_diagram} The phase diagram for La$_{2-x}$Sr$_{x}$CuO$_{4}$ (LSCO) is plotted as a function of hole doping.  For low hole dopings, LSCO is an insulating antiferromagnetic.  At higher concentrations high-temperature superconductivity is present.}
\end{figure}

	The simplest high-temperature superconductor is based on monolayered La$_{2}$CuO$_{4}$.  The phase diagram is illustrated in Fig. \ref{phase_diagram}.  The phase diagram is generic amongst all hole doped cuprates and La$_{2}$CuO$_{4}$ doped with Strontium is chosen as an illustrative example.  The hole doped phase diagram is similar to that of electron doping, though the later class of materials will not be described in this review. La$_{2}$CuO$_{4}$ can be driven into the superconducting phase by replacing La with Sr (La$_{2-x}$Sr$_{x}$CuO$_{4}$) or by doping with excess Oxygens to form La$_{2}$CuO$_{4+y}$. La$_{2}$CuO$_{4+y}$ allows the opportunity to study superconductivity in a stoichiometric crystal, free of structural defects.  Superconductivity appears at hole dopings above 5.5\%, reaching a maximum T$_{c}$ of $\sim$ 40 K at approximately 15\% hole doping.  The monolayer cuprates have been investigated in great detail throughout the underdoped phase diagram with much attention focused on characterizing the transition from an antiferromagnetic insulator to a superconductor.  The other commonly studied cuprate is the bilayer YBa$_{2}$Cu$_{3}$O$_{6}$ system which has a maximum T$_{c}$ of 93 K.  This material has two weakly coupled CuO$_{2}$ planes in each unit cell and is doped into the superconducting phase by adding oxygen atoms into the CuO chains in the basal plane of the unit cell.  The presence of chains means that for particular oxygen concentrations, the oxygen atoms form a stochiometric superlattice.~\cite{Liang00:336,Andersen99:317}
This again allows the spin fluctuations to be investigated in the absence of structural disorder for particular oxygen concentrations throughout the phase diagram.  The ability to dope holes while maintaining a well ordered structure has proven important in separating the effects of charge doping from those caused by the introduction of structural disorder into the crystal.

	We present a review of the magnetic properties of the hole-doped cuprate superconductors using magnetic neutron scattering.  Emphasis is placed on the underdoped side of the phase diagram for concentrations below where T$_{c}$ is maximum.  The paper is divided into four sections dealing with various aspects related to the superconductivity with emphasis on how the superconducting and magnetic order parameters are related.  The first section examines the common magnetic dispersion measured throughout the cuprate phase diagram and shows a remarkably common underlying spin response amongst the hole-doped cuprates.  The second section deals with the spin structure and how it changes on continuously doping holes from a N\'{e}el-ordered state to a high-temperature superconductor and how the magnetism is related to the presence of impurities.  The third section deals with competing magnetic and superconducting order parameters as investigated through the tuning of structural order and the application of large magnetic fields.  The fourth and final section of this review examines various common trends in the high-temperature cuprate phase diagram.

\section{Universal dispersion in insulating and underdoped phases}

	Neutron inelastic scattering has played an important role in characterizing the cuprates, and in particular, in understanding the insulating and underdoped regions of the phase diagram.   The low-energy magnetic fluctuations have been well investigated using triple-axis spectrometers at thermal and cold reactor sources.  However, the large super-exchange interaction of the Cu$^{2+}$ spins has meant that a full understanding of the high-energy spin fluctuations, and hence the spin interactions, can only be obtained using new instrumentation at spallation neutron sources and accordingly there have been many recent developments towards understanding the spin properties of the cuprates.  This section describes the magnetic dispersion in the monolayer La$_{2-x}$(Sr,Ba)$_{x}$CuO$_{4}$ (LSCO or LBCO) and bilayer YBa$_{2}$Cu$_{3}$O$_{6+x}$ (YBCO) superconductors.  It is divided into two segments dealing with the low-energy fluctuations investigated using cold and thermal neutron sources, and high-energy fluctuations probed using epithermal neutrons from spallation neutron labs.

\begin{figure}[t]
\begin{center}
\includegraphics[width=8.7cm] {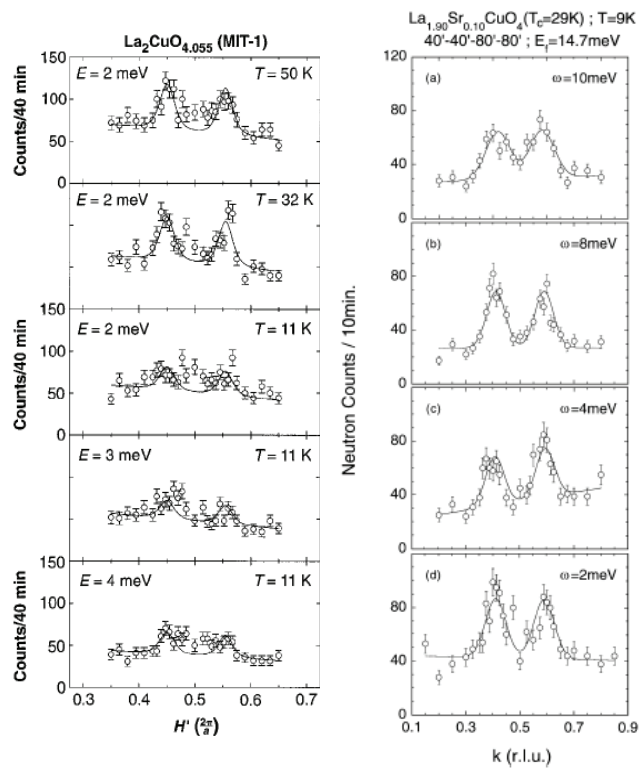}
\end{center}
\caption{\label{incommensurate_fluctuations} Scans through the correlated magnetic scattering at several energy transfers in excess oxygen La$_{2}$CuO$_{4.055}$ and La$_{1.9}$Sr$_{0.1}$CuO$_{4}$.~\cite{Wells97:77,Lee99:69}  The data show incommensurate peaks with little dispersion or broadening with increasing energy transfer.}
\end{figure}

\subsection{Low-energy fluctuations}

	Low-energy fluctuations in insulating LSCO and YBCO show very similar behavior, illustrating a common antiferromagnetic ground state in the insulating cuprates.~\cite{Peters88:37,Tranquada89:40}  Both insulating YBCO and La$_{2}$CuO$_{4}$ are dominated by large nearest neighbor superexchange $J$ $>$ 100 meV and small in- and out-of-plane anisotropies.  The excitations are well understood in terms of linear spin-wave theory. Upon hole doping, the commensurate antiferromagnetic order is destroyed in favor of an incommensurate spin-density wave characterized by magnetic peaks symmetrically displaced away from the ($\pi$,$\pi$) position.  This has been investigated in detail over a broad range in energy transfers for a variety hole concentrations in LSCO between $p$=0.07-0.2.  The magnetic excitations for energy transfers below $\sim$ 20 meV consist of nearly vertical rods extending from the incommensurate positions.~\cite{Lee99:69,Hiraka01:70,Mason93:71}  Further work on excess oxygen doped La$_{2}$CuO$_{4+y}$ has shown that the low-energy magnetic incommensurate scattering is present in the case of excess oxygen doping where effects due to structural disorder may be removed.~\cite{Wells97:77} The incommensurate scattering at low-energies is illustrated in Fig. \ref{incommensurate_fluctuations}, which shows data from excess oxygen La$_{2}$CuO$_{4.055}$ and La$_{1.9}$Sr$_{0.1}$CuO$_{4}$ at several energy transfers. This magnetic incommensurate scattering is directly coupled with the superconductivity as illustrated by two key experimental observations.  First, the value for the incommensurate wave vector defining the displacement of the magnetic scattering away from the ($\pi$,$\pi$) position scales with the superconducting transition temperature.  This property will be discussed in detail later.   Second, for a broad range of hole dopings, the low-energy magnetic response is suppressed for temperatures below the superconducting transition temperature, indicating the formation of a spin gap.  These two points suggest that the incommensurate magnetic scattering is directly associated with high-temperature superconductivity.  

	The temperature dependence of the magnetic scattering at finite energy transfers in the normal state has been shown to follow a simple $\omega/T$ scaling law over a very broad range of hole-dopings extending from the spin-glass to the optimally doped regions.~\cite{Keimer92:46,Keimer91:67,Aeppli97:278}  Scaling has also been observed in La$_{2}$Cu$_{1-x}$Li$_{x}$O$_{4}$ implying that this property is common among the doped monolayer cuprates.~\cite{Bao03:91} The only  assumption behind the presence of scaling is that the dominant energy scale is defined by the temperature and naturally this assumption is expected to break down at energies comparable with the anisotropic exchange terms.  At low-energies, a breakdown of scaling has been observed and thought to be the result of the out of plane anisotropy in the spin exchange.~\cite{Matusda93:5}

\begin{figure}[t]
\begin{center}
\includegraphics[width=7.0cm] {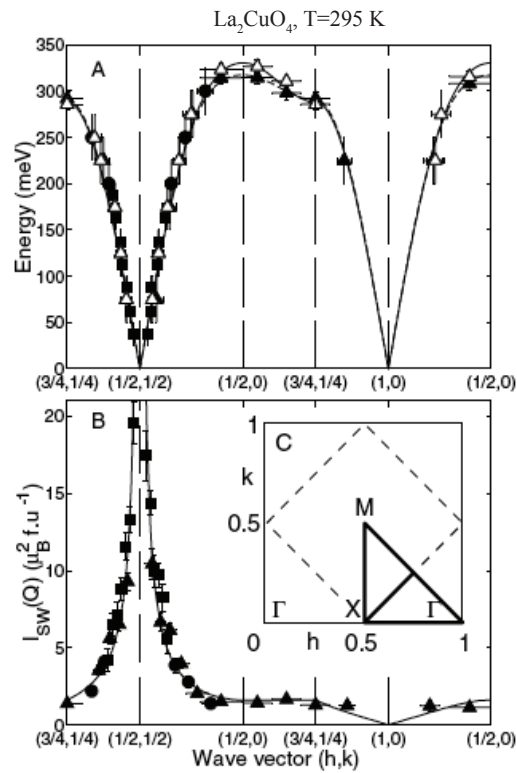}
\end{center}
\caption{\label{insulator} The full dispersion of the magnetic excitations in insulating La$_{2}$CuO$_{4}$ across the entire Brillouin zone.  The anisotropy around the zone boundary is indicative of strong next nearest neighbor interactions.  The data is taken from Coldea \textit{et al.}~\cite{Coldea01:86} }
\end{figure}

	The low-energy excitations in the bilayer YBCO$_{6+x}$ system have been studied in most detail for strongly superconducting samples.~\cite{Fong00:61,Dai01:63,Stock04:69}  Below T$_{c}$, the excitation spectrum is dominated by a strong commensurate peak in the neutron cross-section.  At optimal doping the peak is centered at 41 meV and was originally thought to define a spin-gap and to be associated with the pairing boson for superconductivity or with the presence of a pseudogap in the spin dynamics.~\cite{Dai00:406,Timusk99:62}  Recent studies on underdoped YBCO have found that there is indeed significant magnetic scattering below the resonance energy with a similar incommensurate and anisotropic lineshape and spectral weight to that measured in the monolayer LSCO.~\cite{Mignod92:180,Mook00:404,Bourges00:288,Hinkov04:430,Stock04:69}  It has also been demonstrated through a number of experiments by many groups that the incommensurability in YBCO$_{6+x}$ scales with the superconducting transition temperature in a similar manner to that carefully measured in the monolayer cuprates,~\cite{Balatsky99:82}  confirming that indeed the magnetic fluctuations are directly coupled with the superconductivity.  The anisotropy of the momentum lineshape at low energies where the incommensurate peaks are predominately displaced along the [100] directions implies that an explanation in terms of a simple conventional Fermi-surface nesting is unlikely and therefore models involving stripes or spiral spin phases may be more appropriate.~\cite{Hasselmann04:69}  This point is further emphasized through the study of Lee \textit{et al.} on excess oxygen stage-4 LaCuO$_{4+y}$ (T$_{c}$=42 K) who found the magnetic peaks incommensurate with the lattice and hence not precisely aligned with the Cu-O-Cu tetragonal directions.~\cite{Lee99:60}   However, we note that recent calculations have shown that models based on transitions from a truncated Fermi surface may indeed explain the anisotropy observed in the cuprates.~\cite{Bascones05:xx}  Further work and experiments are required to distinguish between a quasiparticle or spin based model. 

	The combined results on excess oxygen La$_{2}$CuO$_{4+y}$, YBCO, and LSCO confirm that the incommensurate scattering is a common feature to all cuprates and is not unique to LSCO nor the monolayer cuprates.   The magnetic fluctuations in the very underdoped region of the YBCO$_{6+x}$ diagram have recently been studied in detail due to recent advances in materials preparation.~\cite{Liang02:383, Peets02:15}  The results showed a similar energy lineshape in the inelastic channel to that observed in very underdoped LSCO.~\cite{Stock06:73}  The temperature dependence of the low-energy fluctuations in doped YBCO$_{6+x}$ is also well described by an $\omega/T$ law as is the case in LSCO.~\cite{Birgeneau92:87}  It is interesting to note that scaling is followed in both the YBCO$_{6+x}$ and LSCO cuprates over a very broad range of hole dopings implying that the origin of scaling is not simply the close proximity of a quantum critical point.

\begin{figure}[t]
\begin{center}
\includegraphics[width=8.5cm] {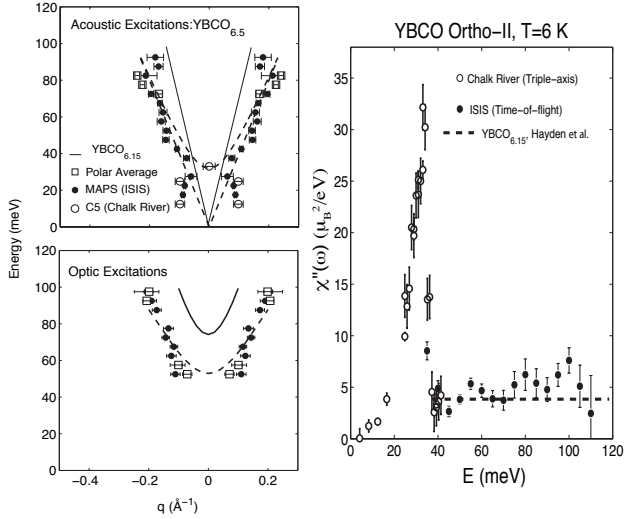}
\end{center}
\caption{\label{YBCO_dispersion}  The magnetic dispersion and integrated weight for YBCO$_{6.5}$ is plotted as a function of energy transfer.  The dotted line in the right hand panel represents the spectral weight expected for the insulating cuprates.  The data is taken from Stock \textit{et al.}~\cite{Stock05:71}}
\end{figure}

\subsection{High-energy spin fluctuations}

	Measurements of the complete dispersion across the entire Brillouin zone provides a direct measure of the spin Hamiltonian.  While measuring the low-energy spin response and the velocity of spin-waves gives the strength of nearest neighbor interactions, it is only a complete measure of the dispersion near the magnetic zone boundary that provides definitive information on the strength of higher order interactions.  Measurements of the low-energy spin response on single layer La$_{2}$CuO$_{4}$ and bilayer YBa$_{2}$Cu$_{3}$O$_{6}$ were first conducted using thermal triple-axis spectrometers but were limited in the energy range which could be covered.

	High energy measurements on LSCO and YBCO were first conducted on hot reactor neutron sources.~\cite{Reznik96:53}  These measurements were able to observe successfully the spin-wave velocity and to observe the presence of an optic mode in the bilayer YBCO system. Definitive studies of the spin-wave velocity were later made use spallation neutrons which allowed measurements in excess of 100 meV in energy transfer to be conducted.  Early measurements by Hayden \textit{et al.} at ISIS on La$_{2}$CuO$_{4}$ and YBa$_{2}$Cu$_{3}$O$_{6}$ found nearest neighbor exchange constants of about 125 meV, consistent with triple-axis work done at low-energies.  However, these early studies were not able to obtain data near the magnetic zone boundary and it was only more recently that Coldea \textit{et al.} were able to measure the dispersion in La$_{2}$CuO$_{4}$ across the entire Brillouin zone.~\cite{Coldea01:86}  These experiments found the dispersion around the magnetic zone boundary to be anisotropic and characteristic of a ferromagnetic next-nearest exchange.  An alternate explanation in terms of a ring exchange term was postulated.  The spin-wave dispersion and intensity as a function of momentum transfer throughout the magnetic zone is shown in Fig. \ref{insulator}.  The solid lines are calculations based on linear spin-wave theory with nearest neighbor exchange interactions of J=112 $\pm$ 4 and next-nearest J$'$=-11 $\pm$ 3 meV.

\begin{figure}[t]
\begin{center}
\includegraphics[width=8.5cm] {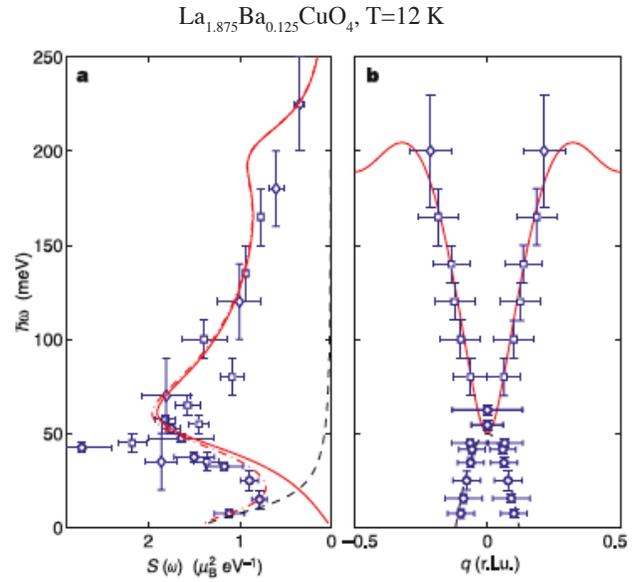}
\end{center}
\caption{\label{LBCO_dispersion} The magnetic dispersion and integrated weight for La$_{0.875}$Ba$_{0.125}$CuO$_{4}$ is plotted as a function of energy transfer. The data is taken from Tranquada \textit{et al.}~\cite{Tranquada04:429}}
\end{figure}

	Measurements of the high-energy dispersion in the superconducting state were first conducted by Hayden \textit{et al.} with later experiments studying the dispersion in YBCO by Arai \textit{et al.}~\cite{Hayden96:54, Hayden96:76, Arai99:83}  Further work on LBCO, YBCO, and LSCO was conducted using improved instrumentation on the MAPS spectrometer at the  ISIS spallation source.~\cite{Stock05:71,Hayden04:429,Tranquada04:429,Christensen04:93}  The dispersion and integrated intensity as a function of energy transfer is plotted in Fig. \ref{YBCO_dispersion} for superconducting Ortho-II YBCO$_{6.5}$ and in Fig. \ref{LBCO_dispersion} for LBCO.  The magnetic response in YBCO consists of both optic and acoustic modes resulting from the fact that it is a bilayer system with two Cu$^{2+}$ ions in each unit cell.  In our discussion we will focus on the acoustic mode only.  The data for the acoustic scattering show two energy scales, a low energy region where the dispersion is characterized by incommensurate scattering with an almost infinite slope and a high-energy scale where the excitations disperse with increasing energy. Measurements on optimally doped YBCO$_{6.85}$ have been conducted by Pailhes \textit{et al.}~\cite{Pailhes04:93} and show a very similar dispersion to measurements conducted in the underdoped region of the phase diagram.

	The experiments focused on the high-energy excitations have provided much new information on the spin spectrum and helped to clarify previous apparent differences between the bilayer YBCO and monolayer LSCO systems.  The measurements conducted on LBCO and LSCO show that the low-energy incommensurate scattering do meet at a resonance energy which in LBCO is at about 50 meV.  This is larger than the energy of the resonance peak in the YBCO cuprates and shows that the resonance peak is not directly related to the pairing boson in the cuprates as it does not scale with the critical temperature.  The resonance also has too little spectral weight to be associated with the pairing boson as discussed by Kee \textit{et al.}, this is illustrated in Fig. \ref{YBCO_dispersion} which shows that the resonance only makes up a small fraction of the total spectral weight.~\cite{Kee02:88}

	These measurements conducted on several different cuprates over a broad range of superconducting hole dopings show a remarkably consistent dispersion with two distinct energy scales and the resonance defining a cross-over point (Figs. \ref{YBCO_dispersion} and \ref{LBCO_dispersion}). The low-energies are characterized by incommensurate scattering with an anisotropic lineshape in momentum which meet at the resonance energy.  At higher energies, the scattering disperses in a manner similar to that measured in the insulating cuprates with a much reduced spin-wave velocity indicative of a significant weakening of the nearest neighbor super-exchange. Whether or not the momentum lineshape at high-energies is closer to a circle or square is currently not resolved.  However, Tranquada \textit{et al.}~\cite{Tranquada04:429} and Stock \textit{et al.}~\cite{Stock05:71} have shown that the spectral weight at high-energy transfers (Figs. \ref{YBCO_dispersion} and \ref{LBCO_dispersion}) is similar to that measured in the insulator and therefore the high-energy spin excitations appears to be more consistent with spin-wave type excitations.  The recent measurements on the high-energy dispersion of the cuprates have revealed a general dispersion which is common among the monolayer and bilayers systems and possibly to all cuprates.  It will be important for new theories to explain and model this and to explain its difference from the behavior in the insulator. 

\section{Discovery of diagonal spin density modulation}

In the previous section the underlying spin dispersion was discussed throughout the underdoped region of the phase diagram.  The magnetic spectrum in the superconducting state is clearly very different to that of the antiferromagnetic insulator.  The incommensurability, which defines the low-energy spectrum, is directly related to the superconductivity and scales simply with the doping.  In this section, the transition from an antiferromagnetic insulator to a superconductor with incommensurate magnetic fluctuations is discussed in detail and the origin of the incommensurate ordering is examined through impurity doping. 

The linear relation $\delta=x$ in the underdoped superconducting (SC) phase~\cite{Yamada98:57} and the sharp incommensurate (IC) elastic peaks observed in the vicinity of $x$=0.12~\cite{Suzuki99:59} triggered a systematic exploration of spin correlations in the spin-glass (SG) phase. How does the incommensurability disappear towards the undoped insulating phase? Does the sharp IC elastic peaks exist in the wider doping region down to the SG phase?   

\begin{figure}[t]
\begin{center}
\includegraphics[width=8.5cm] {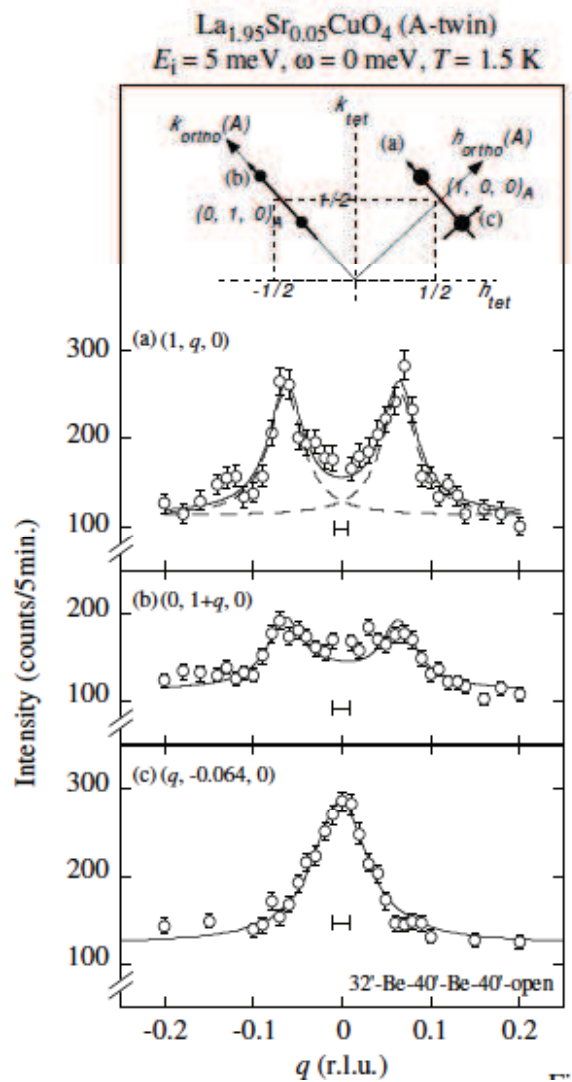}
\end{center}
\caption{\label{boundary_1} Elastic peak profiles of the scans through (100) along (a) the b$^{*}$ and (b) a$^{*}$ axes for the domain A in orthorhombic notation. Scan trajectories are illustrated in the inset. The small horizontal bars indicate the instrumental resolution full width. The data is taken from Wakimoto \textit{et al.}~\cite{Wakimoto00:61}}
\end{figure}

	In order to answer these questions Wakimoto \textit{et al.} revisited the spin correlations in the SG phase near the SC-SG boundary where they discovered a new type of spin density modulation by elastic scattering.~\cite{Wakimoto99:60} We call it a diagonal spin density modulation (D-SDM) because of the IC peak positions are rotated by 45$^{\circ}$ in reciprocal space about ($\pi$,$\pi$) from those observed in the SC phase. Therefore, the modulation runs along the direction diagonal to the Cu-O bonds in CuO$_{2}$ planes, in contrast to the direction in the SC phase which is parallel/perpendicular to the Cu-O bonds. For the latter, we call the modulation in the SC phase a parallel spin density modulation (P-SDM) to distinguish the two types of spin modulation. 

\begin{figure}[t]
\begin{center}
\includegraphics[width=8.5cm] {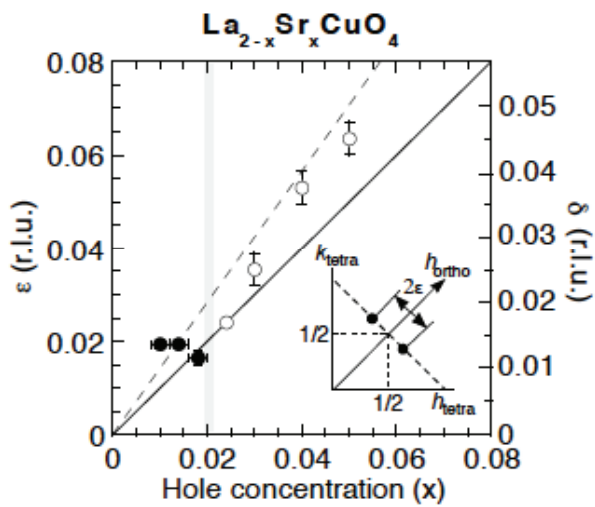}
\end{center}
\caption{\label{boundary_2} Hole concentration $x$ dependence of the splitting of the incommensurate peaks from Matsuda \textit{et al.} ~\cite{Matsuda02:65} The inset shows the configuration of the incommensurate peaks in the D-SDM phase. $\epsilon$=2$\sqrt \delta$ where $\delta$ is defined in tetragonal units. The solid and broken lines correspond to $\epsilon$=x and $\delta$ = x, respectively.}
\end{figure}

One more important discovery by Wakimoto {\it et al.}\cite{Wakimoto00:61} was the one-dimensionality of the D-SDM.  In general, due to the orthorhombic symmetry, crystals contain multiple domains, however, they were fortunate to study a crystal with just two domains, A and B.  A single pair of IC peaks was observed, and it was possible to uniquely associate them with domain A.  As shown in Fig. \ref{boundary_1}, it turned out that the direction of the modulation is the b-axis, which is parallel to the buckling direction of the CuO$_6$ octahedra

\begin{figure}[t]
\begin{center}
\includegraphics[width=8.5cm] {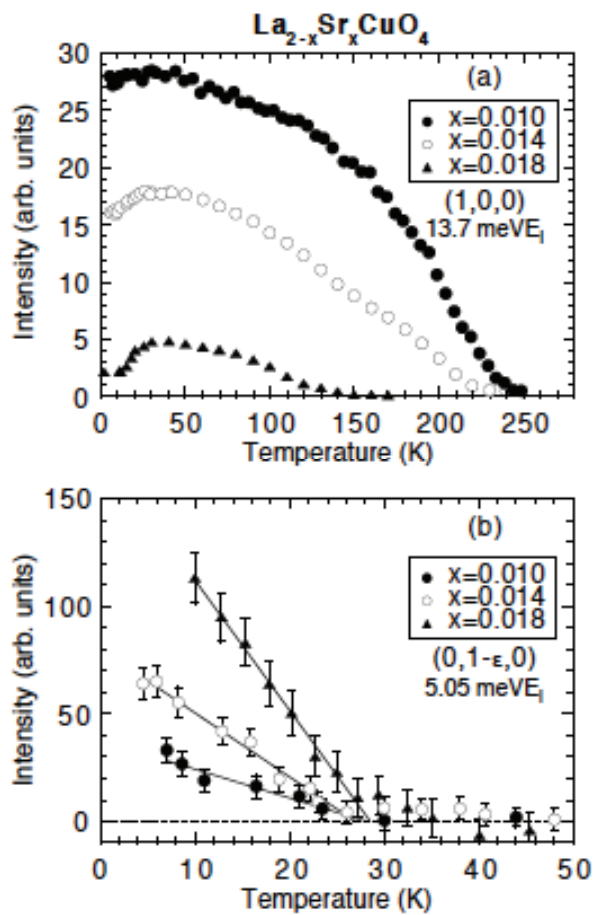}
\end{center} 
\caption{\label{boundary_3} Temperature dependence of the (100) magnetic Bragg intensity (a) and the magnetic intensity at the diagonal incommensurate position (0,1-$\epsilon$,0) (b) in La$_{2-x}$Sr$_{x}$CuO$_{4}$ (x=0.01, 0.014, and 0.018). The solid lines are the results of fits to a linear function. Background intensities measured at a high temperature have been subtracted in (b). The data is taken from Matsuda \textit{et al.}~\cite{Matsuda02:65}}
\end{figure}

	Following these experimental results, Matsuda \textit{et al.}~\cite{Matsuda00:62} performed a systematic neutron scattering study on the doping and temperature dependences of the D-SDM. They observed that the IC elastic peaks appear at low temperatures throughout the SG phase. From scans along the [001] direction which determines the L-dependence of the peak intensities, they further confirmed that the spin correlations are predominantly confined within the CuO$_{2}$ planes, with the Cu$^{2+}$ spins weakly coupled between nearest-neighbor planes. 

\begin{figure}[t]
\begin{center}
\includegraphics[width=7.0cm] {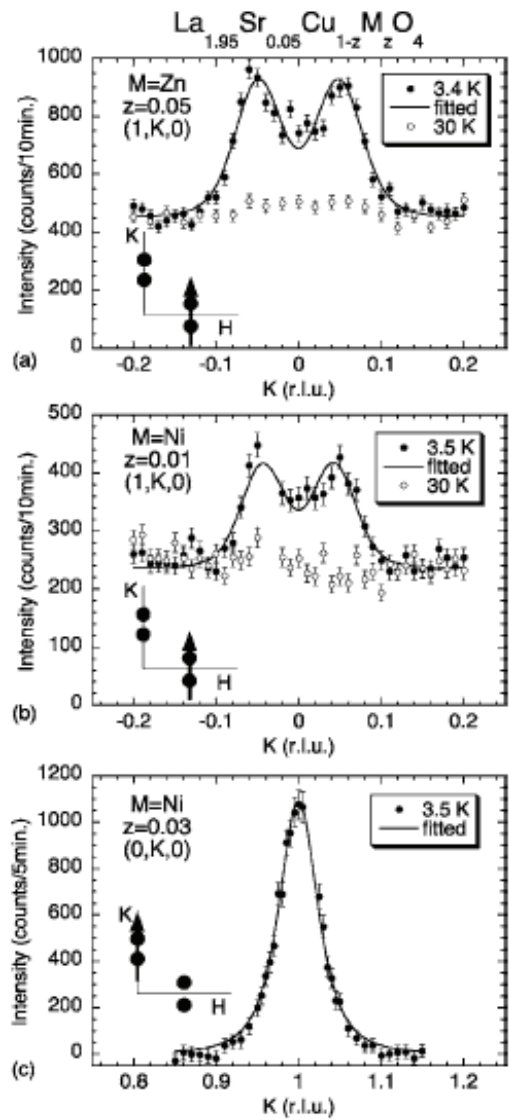}
\end{center}
\caption{\label{boundary_4} Elastic scans along (1,K,0) in La$_{1.95}$Sr$_{0.05}$Cu$_{0.95}$Zn$_{0.05}$O$_{4}$ (a) and La$_{1.95}$Sr$_{0.05}$Cu$_{0.99}$Ni$_{0.01}$O$_{4}$ (b) and along (0,K,0) in La$_{1.95}$Sr$_{0.05}$Cu$_{0.97}$Ni$_{0.03}$O$_{4}$ (c) taken by Matsuda \textit{et al.}~\cite{Matsuda06:xx}. The solid lines are the results of fits of a convolution of the resolution function with 2D squared Lorentzians. Background intensities measured at a high temperature have been subtracted in (c).}
\end{figure}

	Interestingly, the observed incommensurability in the SG phase approximately maps onto the linear relation between $\delta$ and $x$ observed in the SC phase. Based on a simple charge stripe model, this doping dependence suggests that the charge density per unit length of a stripe is almost constant throughout the phase diagram, even when the modulation rotates away by 45$^{\circ}$ near the SC boundary. However, detailed measurements show that there exists a deviation downwards from the linear relation towards the antiferromagnetic (AFM)  phase. In fact as shown in Fig. \ref{boundary_2}, at the lowest values for $x$ the density deviates towards 1 hole/Cu instead of 0.5 hole/Cu as in the parallel charge stripes.

	How does the D-SDM disappear in the AF phase with x $<$ 0.02? Matsuda \textit{et al.}~\cite{Matsuda02:65} examined  the doped AFM phase by a high resolution neutron scattering measurement. Surprisingly, the D-SDM starts to develop below the N\'{e}el temperature and coexists with the 3D-AFM order. The incommensurability, however, does not depend on $x$. Instead, the IC magnetic peak intensity from the D-SDM phase decreases with decreasing $x$. Furthermore, upon cooling the samples, the IC elastic peaks appear simultaneously with the reduction in the commensurate magnetic Bragg peak intensities (Fig. \ref{boundary_3}).   Therefore, the short ranged D-SDM phase precipitates from the AFM phase at low temperature, which suggests that electronic phase separation of the doped holes occurs so that some regions with hole concentration $p$ $\sim$ 0.02 and the rest with $p$$\sim$ 0 are formed. It is noted that the reduction of the magnetic Bragg peak intensity upon cooling as shown in Fig. \ref{boundary_3} was observed by Endoh \textit{et al.} in LCO nearly 20 years ago.~\cite{Endoh88:37} We now finally understand the origin of such phenomena as the precipitation of the D-SDM in the 3D-AFM phase induced by dilute holes from excess oxygen ions in LCO. 

\begin{figure}[t]
\begin{center}
\includegraphics[width=6.0cm] {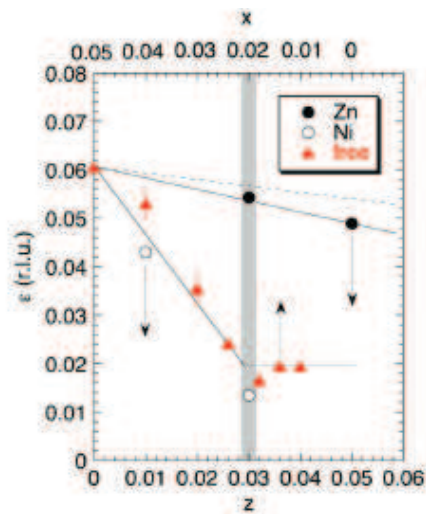}
\end{center}
\caption{\label{boundary_5}  Impurity concentration $z$ dependence of the incommensurability $\epsilon$ in La$_{1.95}$Sr$_{0.05}$Cu$_{1-z}$Zn$_{z}$O$_{4}$ ($z$=0.03 and 0.05) and La$_{1.95}$Sr$_{0.05}$Cu$_{1-z}$Ni$_{z}$O$_{4}$ ($z$=0.01 and 0.03) by Matsuda \textit{et al.}.~\cite{Matsuda06:xx} Hole concentration $x$ dependence of $\epsilon$ in impurity-free La$_{2-x}$Sr$_{x}$CuO$_{4}$~\cite{Matsuda00:62} is also shown. The solid lines are visual guides. The broken line corresponds to $z$ dependence of $\epsilon$ expected in the spiral model.~\cite{Hasselmann04:69} The thick shaded bar represents the boundary between the long-range antiferromagnetic and spin-glass phases in La$_{2-x}$Sr$_{x}$CuO$_{4}$. The incommensurability is almost zero in La$_{1.95}$Sr$_{0.05}$Cu$_{0.9}$7Ni$_{0.03}$O$_{4}$.}
\end{figure}

	What is the direct relevance of the diagonal-parallel SDM transition to the superconductivity? In order to answer this question Fujita \textit{et al.}~\cite{Fujita02:65} prepared single crystals of LSCO with finely tuned hole concentrations near the boundary. Uniform magnetic susceptibility measurements for these single crystals revealed that the SC transition sharply disappears below $x_{c}$ $\sim$ 0.055. Then a detailed neutron scattering study using crystals with $x$ above and below $x_{c}$ finally confirmed the coincidence  between the appearance of the P-SDM and the superconductivity. However, in the SC phase near the boundary the D-SDM and P-SDM with similar incommensurabilities coexist as demonstrated by a circular-scan around ($\pi$,$\pi$). This coexistence suggests a doping-induced first order transition between the diagonal and parallel SDM phases and between the SG and SC phases. It is noted that a $\mu$SR experiment observed the penetration of the SG phase into a wide region of the underdoped SC phase.~\cite{Nied98:80} We speculate that the magnetic signal observed by $\mu$SR corresponds to the elastic magnetic signal from the P-SDM in the underdoped SC phase. 

	What is the microscopic origin of the D-SDM? As in the case of the P-SDM, one possibility to explain the D-SDM is the charge stripe model. The parallel stripe phase in La$_{2-x-y}$Nd$_{y}$Sr$_{x}$CuO$_{4}$ and La$_{2-x}$Ba$_{x}$CuO$_{4}$ is stabilized by the structural distortion, namely, the low temperature tetragonal (LTT) structure, which is strongly coupled to the charge ordering.~\cite{Ichikawa00:85,Fujita02:88} However, in lightly hole-doped LSCO, which has no LTT phase, no evidence has  been found that the charge stripe phase is realized. Most importantly, the structural distortion, originating from the charge ordering, has not been observed, although this non-observation can be due to disorder in the periodicity and direction of the stripe. On the other hand, Hasselman \textit{et al.}~\cite{Hasselmann01:56, Hasselmann04:69} reported that the D-SDM may be explained using the spiral spin model, originating from the magnetic frustration around the localized hole spins. In this model, the IC magnetic peaks are purely magnetic in origin and have nothing to do with charge ordering. 

	Matsuda \textit{et al.}~\cite{Matsuda06:xx} studied the effects of impurities to differentiate between different models for  the D-SDM. The results show that Zn doping reduces the incommensurability just slightly. On the other hand, Ni doping quickly destroys the incommensurability and restores the N\'{e}el ordering, indicating a strong effect on hole localization. This suggests that Ni is doped as Ni$^{3+}$ or as Ni$^{2+}$ with a hole forming a strongly bound state. Therefore, Ni doping reduces the number of mobile or hopping Zhang-Rice (ZR) singlet states around the Cu spins by creating localized hole sites near the doped Ni. Then the concentration of the mobile ZR singlet $x_{eff}$ can be described by the difference between the number of holes and doped Ni ions. In fact, the incommensurability of the D-SDM for the Ni doped samples can be summarized by $x_{eff}$ as shown in Fig. \ref{boundary_5}. Furthermore, the onset temperatures of the D-SDM in Ni doped samples correspond to those of Ni free samples with hole concentration $x_{eff}$. This means that the incommensurability in this system is controlled by the number of mobile ZR singlets or mobile holes rather than the number of localized holes around Ni. Furthermore, it should be noted that due to the strong localization the effective spin value near the doped Ni is expected to be 1/2, the same as the surrounding Cu spins. Hence the magnetic effect of Ni doping should be smaller than that of Zn in contradiction to the experimental result. Therefore, the observed impurity effect on the spin density modulation, particularly the Ni effect favors the charge stripe model based on charge segregation by mobile holes rather than the spiral spin model based on localized hole spins. It is noted that the strong localization effect around Ni impurities is also observed in the lightly doped AFM phase, where the N\'{e}el temperature quickly recovers and the precipitation of the D-SDM is suppressed by Ni doping.~\cite{Hiraka05:74}    

	The comparison between doping holes and impurities directly on the Cu$^{2+}$ sites reveals a strong connection between antiferromagnetism and superconductivity.  These competing order parameters are further discussed in the next section by tuning the structural order through annealing and the superconducting order parameter directly by applying a magnetic field.  

\section{Competing Superconducting and Magnetic Order}

	As discussed in the previous three sections, the Copper oxide high temperature superconductors exhibit remarkably rich and complicated static and dynamic spin fluctuation phenomena across the entire phase diagram.~\cite{Kastner98:70}  It is particularly striking that in underdoped La$_{2-x}$Sr$_{x}$CuO$_{4}$ and related compounds static incommensurate magnetic order and superconductivity coexist.  It is found further that for optimally doped La$_{2}$CuO$_{4.11}$, which is a stoichiometric ordered, single crystal phase without appreciable structural disorder, the superconducting and magnetic phase transitions occur at the same temperature.~\cite{Lee99:69}  It is important to know whether or not these two kinds of order compete or cooperate with one another and whether they coexist microscopically or form spatially separate phases.

\begin{figure}[t]
\begin{center}
\includegraphics[width=8.0cm] {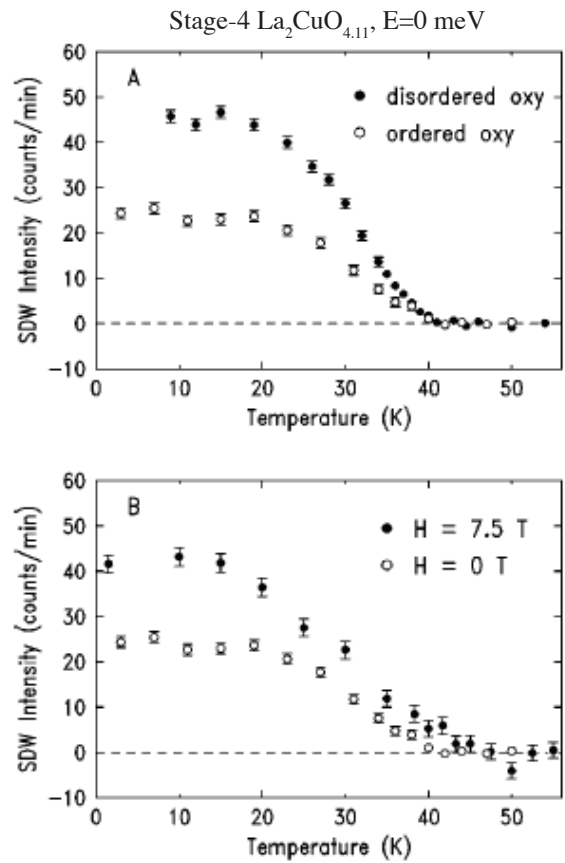}
\end{center}
\caption{\label{disorder_field} Temperature dependence of the elastic SDW peak. (A) The peak intensity in the order-oxygen state versus the disordered-oxygen state. (B) The peak intensity in an applied magnetic field of 7.5 T as a function of temperature.  The data is taken from Lee \textit{et al.}~\cite{Lee04:69}}
\end{figure}

     	As is well known, the superconducting order may be diminished by either the introduction of structural disorder or by the application of a magnetic field which in type II superconductors in turn introduces vortices into the superconductor.  Recently, a number of neutron scattering experiments have been carried out which probe the superconductivity and magnetic correlations in the presence of structural disorder and/or magnetic fields applied perpendicular to the CuO$_{2}$ planes.~\cite{Lee04:69,Katano00:62,Lake01:291,Lake02:299,Khaykovich02:66,Matsuda02:66,Wakimoto03:67,Khaykovich05:71,Lake05:4}  These experiments have in turn stimulated important theoretical developments, most notably by Demler \textit{et al.},~\cite{Demler01:87,Zhang02:66,Arovas97:79,Kivelson01:98} which have elucidated the nature of the competition between the incommensurate magnetic order and the superconductivity.  We will discuss the predictions of these theories in the context of the experiments.  We should emphasize that in this section we will not attempt a comprehensive review of this growing subfield of high temperature superconductivity but rather we shall give a few illustrative examples to evince the basic phenomenology and the essential theoretical ideas.

      	The first experiments to probe these effects were the studies by Yamada and coworkers of La$_{1.88}$Sr$_{0.12}$CuO$_{4}$ (T$_{c}$=12 K) in a magnetic field.~\cite{Katano00:62}  However, since microscopic structural disorder is a feature of nearly all high temperature superconductors, it is convenient to review first the work of Y.S. Lee \textit{et al.}~\cite{Lee04:69} probing the effects of disorder.  Generally, it is difficult to isolate the effects of disorder on any phase transition, especially one with competing order parameters.  However, as discovered by Lee \textit{et al.},~\cite{Lee04:69} stage-4 La$_{2}$CuO$_{4.11}$ is a system in which the intercalated oxygen atoms can be progressively disordered by varying the annealing conditions.  Specifically, for samples annealed at 320 K or lower the intercalated oxygen atoms order three dimensionally so that the material is a true stoichiometric single crystal.  In such samples the superconducting transition temperature is $\sim$ 42 K (onset).  Incommensurate magnetic order with incommensurability of $\sim$ 1/8 occurs at precisely the same temperature.~\cite{Lee99:60}  By annealing the sample at temperatures of 330 K or higher and then quenching to low temperatures the oxygen lattice can be completely disordered.  Measurements of the diamagnetic susceptibility show that for an anneal at 360 K and subsequent quench to 9 K  T$_{c}$ is reduced by 5 K to 37 K.  We show in Fig. \ref{disorder_field} A the intensity of the spin density wave (SDW) peak as a function of temperature in the oxygen ordered state versus the oxygen disordered state.   Fig \ref{disorder_field} B shows data at $H=0$ T and $H=$7.5 T.  We will discuss these latter data later in this section.  As is evident from Fig. \ref{disorder_field} A the effect of disordering the intercalant oxygen atoms is to enhance the magnetic intensity by nearly a factor of 2 while reducing T$_{c}$ by about 5 K.  None of the magnetic phase transition temperature, the magnetic wave vector or the magnetic correlation length is measurably affected.   These results indicate that the superconductivity and the SDW order compete and that microscopic structural disorder favors the magnetic order over the superconductivity.  The sensitivity of the SDW amplitude to structural disorder also explains the variability of the magnetic intensity in different samples of nominally the same composition.~\cite{Lee99:60,Lee04:69,Katano00:62,Lake01:291,Lake02:299,Khaykovich02:66}

\begin{figure}[t]
\begin{center}
\includegraphics[width=8.0cm] {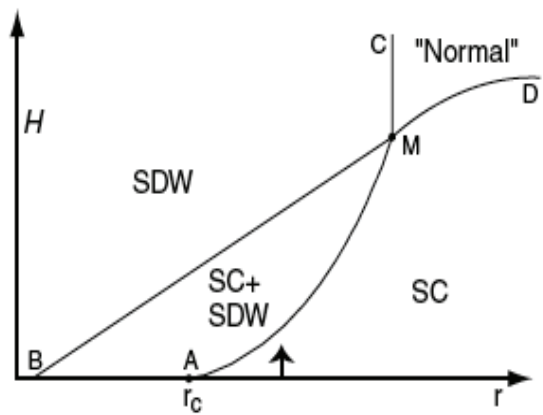}
\end{center}
\caption{\label{Demler_phase} A schematic of the phase diagram predicted by the Ginzburg-Landau theory of Demler \textit{et al.}~\cite{Demler01:87}.  The theory offers a complete description of the  SC to SC+SDW transition at small H.}
\end{figure}

    	As noted above, the first experiments to probe the effects of a magnetic field on the magnetic and superconducting order were those of Katano \textit{et al.}~\cite{Katano00:62}  on La$_{1.88}$Sr$_{0.12}$CuO$_{4}$ (T$_{c}$=12 K). They found that a field of 10 T applied perpendicular to the CuO$_{2}$ planes severely depressed Tc while the low temperature magnetic intensity increased by as much as 50\% suggesting that the superconductivity and SDW order compete.  These experiments were followed by a comprehensive study by Lake \textit{et al.}~\cite{Lake01:291}  of the low energy spin dynamics in the optimally doped material La$_{1.84}$Sr$_{0.16}$CuO$_{4}$ (T$_{c}$=38.5 K) at $H=$0 T and $H=$7.5 T. This study showed, in addition to the expected suppression of T$_{c}$, that a field of 7.5 T created new spin excitations in the superconducting spin gap.  They related this explicitly to the antiferromagnetism in the cores of the vortices and they then made the prescient deduction that the spins in the vortices are correlated over a variety of length scales from the atomic to the mesoscopic.

  	Stimulated by these experiments, Demler \textit{et al.}~\cite{Demler01:87,Zhang02:66}  proposed a Ginzburg-Landau model for coupled SC and SDW order. Their model built on earlier work by a number of authors.~\cite{Arovas97:79,Kivelson01:98, Kivelson03:75}  They hypothesized that these data could be understood by assuming that the superconductor (SC) was in the vicinity of a bulk quantum phase transition to a state with microscopic coexistence of SC and SDW order.  This led them to predict the phase diagram shown in Fig. \ref{Demler_phase}.  They further predicted that in the state with both SC and SW order, application of a magnetic field should cause the magnetic intensity to grow like $\Delta I$ $\sim$ $H/H_{c2} \ln (a H_{c2}/H)$.

\begin{figure}[t]
\begin{center}
\includegraphics[width=8.0cm] {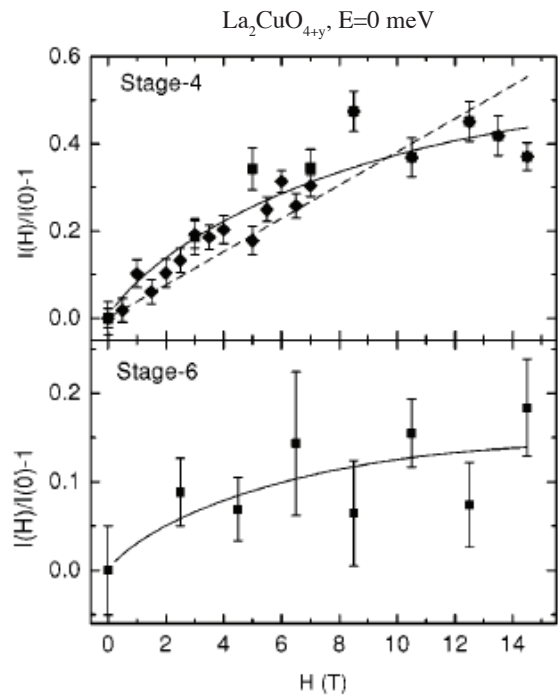}
\end{center}
\caption{\label{intensity_H} Field dependence of the elastic SDW peak intensity.  The solid lines are fits to the formula $\Delta I$ $\sim$ $H/H_{c2} \ln (a H_{c2}/H)$.  The data is taken from Khaykovich \textit{et al.}~\cite{Khaykovich02:66}}
\end{figure}

    	Essentially simultaneous with this theoretical development, Lake \textit{et al.}~\cite{Lake02:299} and Khaykovich \textit{et al.}~\cite{Khaykovich02:66} measured the enhancement of the SDW intensity as a function of field in La$_{1.9}$Sr$_{0.1}$CuO$_{4}$ and stage-4 La$_{2}$CuO$_{4.11}$ respectively.  In both cases a significant enhancement of the SDW intensity is found.  Importantly, the systems exhibit resolution limited magnetic diffraction peaks whose wave vectors are identical at high fields (H up to 14 T) to those of the zero field state.  This necessitates that the induced magnetism has the same microscopic origin as that at zero field.  Data in La$_{2}$CuO$_{4.11}$ at $H=$0 T and 7.5 T are shown in Fig. \ref{disorder_field} B.  In both cases the intensity increases linearly with field as opposed to the H$^{2}$ dependence expected from a pure magnetic effect.  Specifically, this singular dependence of the SDW amplitude on the applied field excludes purely magnetic mechanisms for the phenomenon such as a suppression of the fluctuations of the ordered moment by the applied field or an increase of the correlations along the c-axis each of which by symmetry must scale like H$^{2}$.  This result alone dictates that the SDW enhancement must be driven by the vortices whose density scales linearly with H.  In fact, in both systems it is found that over the entire range of fields the intensity follows the Demler \textit{et al.} prediction, $\Delta I$ $\sim$ $H \ln (a H/H_{c2})$ to within the experimental errors.  Data for stage-6 and stage-4 La$_{2}$CuO$_{4+y}$ for fields from 0 to 14 T are shown in Fig. \ref{intensity_H} together with the $H \ln (1/H)$ theoretical prediction.

     	Indirect confirmation of the model is given by the study by Matsuda \textit{et al.}~\cite{Matsuda02:66} of the effects of the application of a magnetic field on the diagonal stripe spin glass phase in lightly doped La$_{2-x}$Sr$_{x}$CuO$_{4}$.  In this case the magnetic intensity is diminished rather than enhanced by the magnetic field as expected for a purely magnetic mechanism.    A similar result is obtained in La$_{1.45}$Nd$_{0.4}$Sr$_{0.15}$CuO$_{4}$~\cite{Wakimoto03:67};  the authors explain this based on the relative volume fraction exhibiting stripe order at zero field.
	
\begin{figure}[t]
\begin{center}
\includegraphics[width=7.5cm] {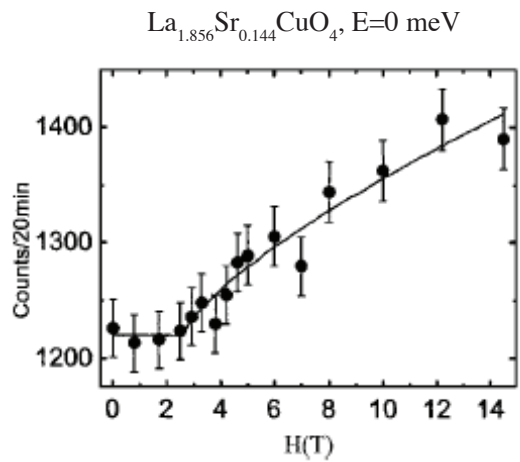}
\end{center}
\caption{\label{LSCO_IvsH} Field dependence of the elastic magnetic scattering corresponding to the incommensurate SDW peak.  The solid line is a fit to the relation $I(H)-I(BG)$= $A (H-H_{c})^{\beta}$.  The data is taken from Khaykovich \textit{et al.}~\cite{Khaykovich05:71}}
\end{figure}

    	The most dramatic prediction of Demler \textit{et al.}~\cite{Demler01:87, Zhang02:66} may be immediately seen from Fig. \ref{Demler_phase}.  For a sample which is in the pure superconducting state but in the immediate vicinity of the quantum critical point (QCP), which is at $H=$0, $r=r_{c}$ in Fig. \ref{Demler_phase}, it should be possible to drive the system from the SC to the SC+SDW state with the application of a magnetic field and the transition should be second order.  Consideration of all of the available neutron scattering data in the La$_{2-x}$Sr$_{x}$CuO$_{4}$ system led Khaykovich \textit{et al.}~\cite{Khaykovich05:71} to conclude that if Fig. \ref{Demler_phase} is correct then the QCP should lie near $x$ $\sim$1/8.  Accordingly, Khaykovich \textit{et al.} studied the magnetic response as a function of magnetic field in a crystal with $x$=0.144, intermediate between the putative QCP and optimal doping.  They confirmed earlier results that at $H=$0 T no static magnetic order is observable.  However at fields above about 3 T a clear SDW signal corresponding to long range incommensurate order is observed.  The detailed temperature dependence of the SDW diffraction signal is shown in Fig.\ref{LSCO_IvsH}.  The intensity is fitted according to a power law $I(H)-I(BG)= A (H-H_{c})^{\beta}$, with $H_{c}$=2.7$\pm$0.8 T and $\beta$=0.36 $\pm$ 0.10.  This experiment would seem to provide convincing evidence for the basic physics of the model of competing SC and SDW order proposed by Demler \textit{et al.}~\cite{Demler01:87,Zhang02:66}  Further studies of both the static and dynamic magnetic behavior as a function of magnetic field around $x$=1/8 are required to complete the empirical picture and to determine definitively whether or not $x$=1/8 is indeed a QCP in the La${2-x}$Sr$_{x}$CuO$_{4}$ system.  

\begin{figure}[tb]
\begin{center}
\includegraphics[width=0.85\columnwidth]{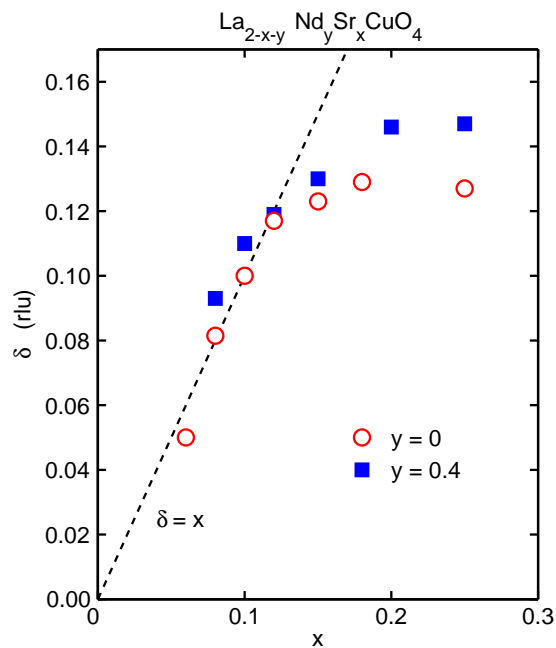}
\end{center}
\caption{Variation of the magnetic incommensurability $\delta$ in \lsco\
with and without Nd codoping.  The open circles are from measurements of
excitations at $\hbar\omega\sim3$~meV ant $T\approx T_c$ in \lsco\ from
Yamada {\it et al.}\cite{Yamada98:57}; filled squares are from elastic
scattering on \lnsco\ from Ichikawa {\it et al.}\cite{Ichikawa00:85}}
\label{fig:e_vs_x}
\end{figure}

\section{Putting the pieces together}

We have noted that the magnetic correlations in under-doped \lsco\ are
compatible with a model in which the spin incommensurability is a result
of the segregation of the doped holes into stripes, forming antiphase
domain walls.\cite{Kivelson03:75,zaan01,sach91,mach89,cast95}  There is direct
evidence for order of this kind in \lnsco\ and \lbco\ (especially at
$x=\frac18$) from diffraction
measurements\cite{Ichikawa00:85,tran95a,Fujita02:88,fuji04}; the charge order has
recently been confirmed by soft--x-ray resonant
diffraction.\cite{abba05}  By segregating the charge carriers, it is
possible for magnetic domains to survive that are similar to those of the parent insulator.  The superexchange interaction should still be operable in
these domains, and the large energy scale for magnetic excitations found
in \lsco\ and \lbco, as discussed in \S2, is consistent with this
picture.  The similar dispersion of spin excitations in underdoped \ybco\
suggests that the same sort of stripe-like correlations are present there as well.

\begin{figure}[tb]
\begin{center}
\includegraphics[width=0.85\columnwidth]{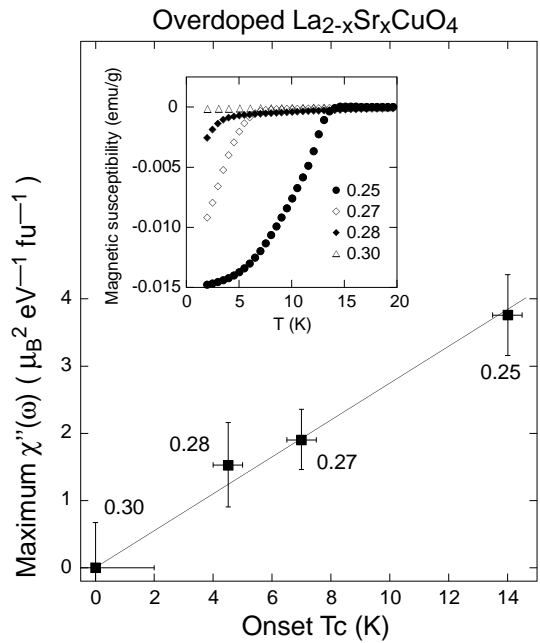}
\end{center}
\caption{Maximum of {\bf q}-integrated $\chi''(\omega)$ at 8~K as a
function of $T_c$ for overdoped \lsco, from Wakimoto {\it et
al.}\cite{waki04}  The solid line is the result of a least-squares fit to
a linear function.  The inset shows the magnetic shielding measured in
10~Oe after cooling in zero field.}
\label{fig:waki}
\end{figure}

The diagonal spin modulation found at light doping in \lsco\ between the
antiferromagnetic and superconducting phases might not be a universal
feature among the cuprates.  Studies of the bilayer system
La$_{2-x}$(Sr,Ca)$_x$CaCu$_2$O$_{6+\delta}$ with $x = 0.1$--0.2
(corresponding to 5--10\%\ hole doping) reveal patches of commensurate
short-range antiferromagnetic order that survive to temperatures
$>100$~K,\cite{ulri02,huck05} even though optical conductivity
measurements demonstrate the presence of mobile holes in the CuO$_2$
planes.\cite{wang03}  The common feature seems to be that doped holes do
not mix well with commensurate antiferromagnetism.  The coexistence of
distinct regions of antiferromagnetic order and of mobile holes in
La$_{2-x}$(Sr,Ca)$_x$CaCu$_2$O$_{6+\delta}$ might reflect substantial
local variations in chemical potential balanced by Coulomb interactions;
in any case, the large electronic disorder in this system is correlated
with a small superconducting volume fraction, as determined by magnetic
susceptibility measurements.\cite{huck05}

Returning to the superconducting regime, the stripe picture is plausible
in the under-doped region, where stripes can maintain a reasonable
separation.  If the stripes all have a similar hole concentration, then
increasing the hole density leads to a rise in the stripe density. 
Experimentally, the magnetic incommensurability, which should be
proportional to the stripe density, increases with doping, as shown in
Fig.~\ref{fig:e_vs_x}, until it saturates above
$x\sim\frac18$.  At $x=\frac18$, the separation between charge stripes is
approximately four lattice spacings; it may be difficult to increase the
stripe density beyond this point, as the stripes would lose their
definition at smaller separations.  

It is informative to consider the behavior at large doping.  Wakimoto and
coworkers\cite{waki04} have studied the magnetic scattering from \lsco\
with $0.25\le x\le0.30$. At low temperatures, a peak in the inelastic
scattering is found at $\sim6$~meV; however, the magnitude of the
imaginary part of the dynamic susceptibility, $\chi''$, decreases steadily
to zero.  As shown in Fig.~\ref{fig:waki}, they have found that the
decrease in the superconducting transition temperature, $T_c$, is correlated with $\chi''$.

\begin{figure}[tb]
\begin{center}
\includegraphics[width=0.95\columnwidth]{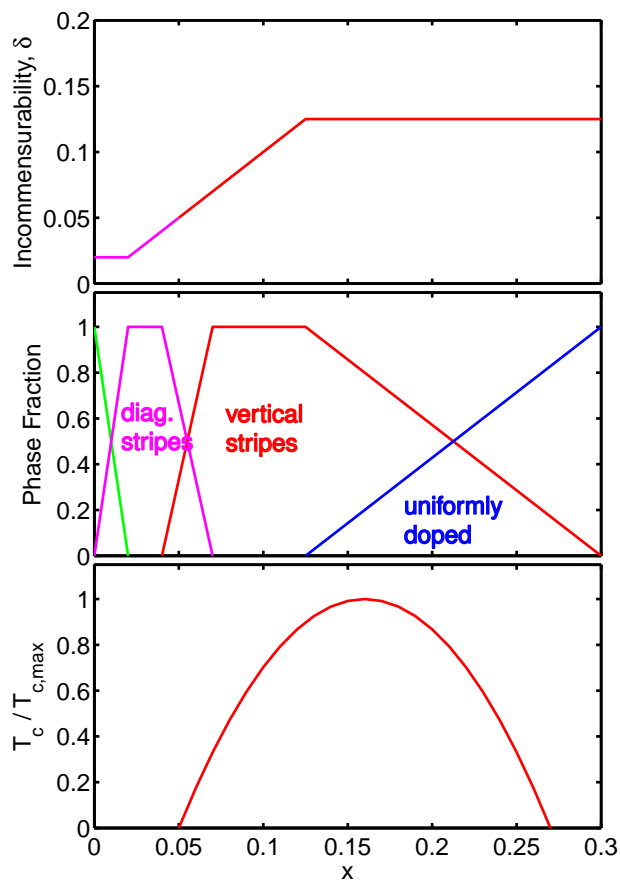}
\end{center}
\caption{Top: schematic diagram of the magnetic incommensurability
measured in \lsco.\cite{Yamada98:57,Matsuda00:62,Matsuda02:65}  Middle: schematic
diagram suggesting regions of existence and coexistence of various
electronic phases; from left, antiferromagnetic order, diagonal stripes,
vertical stripes, uniformly-doped phase.  Bottom: typical curve of $T_c$
(normalized to $T_{c,{\rm max}}$) vs. doping.}
\label{fig:frac}
\end{figure}

What happens to the signal?  One possibility would be for it to move to
higher energy.  The big question is whether the weight remains at an
energy scale of some tens of meV, consistent with fluctuating local
moments, or whether it moves to electronic energy scales, on the order
of eV.  To test this, new measurements at ISIS have looked for magnetic
excitations up to $\sim100$~meV in \lsco\ with $x=0.25$ and
0.30.\cite{waki06}  They demonstrate a drastic reduction in magnetic
signal for
$\hbar\omega>10$~meV, with the signal at lower energies consistent with
the triple-axis work.\cite{waki04}  Thus, the correlation shown in
Fig.~\ref{fig:waki} truly shows that superconductivity and local
antiferromagnetism disappear together.  (Note that it is not practical to
perform similar measurements on \ybco\ because so far it has not been possible to achieve such
high doping levels.)

Muon spin-relaxation studies have shown, in several cuprate systems, that
there is a rapid decrease in the fraction of normal-state charge carriers
that participate in the superfluid density as the hole concentration is
raised above 20\%.\cite{bern01b,uemu01}  There have also been reports of
percolative behavior in over-doped samples based on magnetization studies
of the superconducting state.\cite{wen00}  Such observations have led to
suggestions that over-doped cuprates are characterized by electronic phase separation.\cite{uemu01,wen00}  

\begin{figure}[tb]
\begin{center}
\includegraphics[width=0.95\columnwidth]{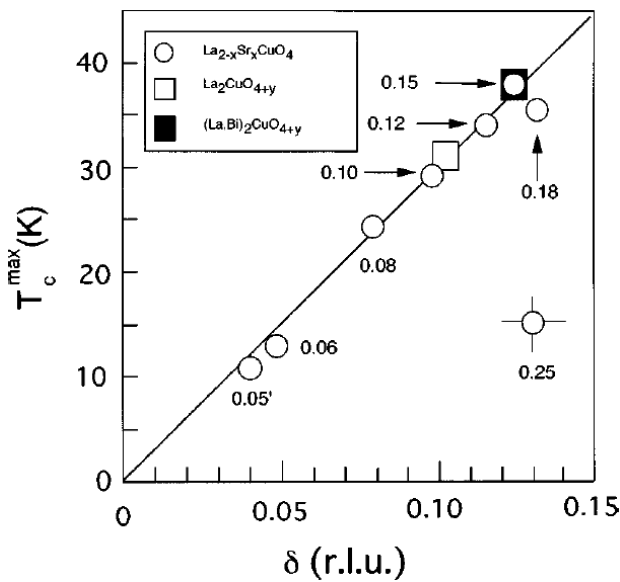}
\end{center}
\caption{A summary of results for $T_c$ vs.\ $\delta$ in \lsco,
La$_2$CuO$_{4+y}$, and related compounds, from Yamada {\it et
al.}\cite{Yamada98:57}}
\label{fig:tc_vs_d}
\end{figure}

If we take the phase-separation scenario seriously, then the fact that the
residual magnetic scattering at high doping is incommensurate suggests
that one of the phases has stripe correlations similar to the $x=\frac18$
phase, with the volume fraction of this phase decreasing with $x$, as
indicated in the middle panel of Fig.~\ref{fig:frac}.  The other phase is
presumably uniformly doped.  The picture, then, is that as one increases
the doping beyond $x\sim\frac18$, it becomes unfavorable to accommodate
the additional holes in stripes; instead, patches of the uniformly-doped
phase grow at the expense of the stripe phase.  The maximum $T_c$ seems
to occur in a mixed phase region dominated by the stripe phase.  

New evidence for stripe correlations in cuprate superconductors comes
from a neutron scattering study of the bond-stretching phonon
mode.\cite{rezn06}  In stripe-ordered samples, a sharp drop in the
dispersion, together with a large increase in energy width, is observed
at ${\bf q} = (0.25,0,0)$, in reciprocal lattice units.  A very similar
anomaly is observed in superconducting \lsco\ and \ybco; however, the
anomaly is absent in nonsuperconducting La$_2$CuO$_4$ and
La$_{1.7}$Sr$_{0.3}$CuO$_4$.  There is a strong circumstantial case that
the anomaly is associated with stripes.  Further support comes from the
absence of such an anomaly in conventional calculations.\cite{bohn03}

Figure~\ref{fig:tc_vs_d} shows the experimental correlation between the
superconducting transition temperature and the magnetic
incommensurability.\cite{Yamada98:57}  The observed trend is consistent with
the theoretical arguments that electronic inhomogeneity (as in the form
of charge  stripes) is important for achieving high-temperature
superconductivity.\cite{kive05a}   The field-induced magnetic order
discussed in \S4 indicates that even static stripes are close in energy
to the superconducting state.

The perspective throughout this article has been one in which the cuprates
are viewed as doped antiferromagnets, with the magnetic correlations
observed by neutron scattering arising from local Cu moments coupled by
superexchange.  In this picture, the carriers tend to segregate to define
the magnetic regions.  There is, of course, an alternative view in which
the magnetic scattering is attributed entirely to the charge carriers. 
It is possible to calculate the dynamic susceptilibility from
conventional formulas for quasiparticles with residual magnetic or
Coulomb interactions.\cite{kao00,norm00,brin99,chub01,onuf02}   Such an
approach is, perhaps, more natural when one comes at the problem from the
over-doped side.  In principle, there should be excitations coming both
from Cu moments and from the carriers.  The challenge for the future is
to learn how these different contributions combine and interact.

\section{Acknowledgment}

Three of us had the chance to enjoy the excitement of working with Gen
Shirane on magnetic neutron scattering studies of the cuprates at the
High Flux Beam Reactor (HFBR).  Even after the HFBR's last cycle, Gen continued
to direct collaborative experiments at international facilities for
another decade.  We have all benefited from Gen's knowledge, experience,
and leadership, and we are deeply grateful for the opportunities that we
had to work with, and learn from, him. 

JMT is supported at Brookhaven by the Office of Science, U.S. Department
of Energy, under Contract No.\ DE-AC02-98CH10886.  This work has
benefited from the U.S.-Japan Cooperative Program on Neutron Scattering.  The work
at Lawrence Berkeley Laboratory is supported by the Office of Basic Energy
Sciences, U.S. Department of Energy under contract number DE-AC03-76SF00098. The work at Johns Hopkins University was funded by the Natural Sciences and Engineering Research Council (NSERC) of Canada and through DMR 0306940. The work at Tohoku University is supported by the Grants-in-Aid for scientific research of Ministry of Education, Culture, Sports, Science and Technology(MEXT) of Japan.


\thebibliography{}


\bibitem{Kastner98:70} M.A. Kastner, R.J. Birgeneau, G. Shirane, and Y. Endoh, Rev. Mod. Phys. {\bf{70}}, 987 (1998).

\bibitem{Kivelson03:75} S. A. Kivelson, I. P. Bindloss, E. Fradkin, V. Oganesyan, J. M. Tranquada, A. Kapitulnik, and C. Howald, Rev. Mod. Phys. {\bf{75}}, 1201 (2003).

\bibitem{Lee06:78} P. A. Lee, N. Nagaosa, and X.-G. Wen, Rev. Mod. Phys. {\bf{78}}, 17 (2006).

\bibitem{Liang00:336} R. Liang, D.A. Bonn, and W.N. Hardy, Physica C {\bf{336}}, 57 (2000).

\bibitem{Andersen99:317} N.H. Andersen, M. von Zimmermann, T. Frello, M. Kall, D. Monster, P.-A. Lindgard, J. Madsen, T. Niemoller, H.F. Poulsen, O. Schmidt, J.R. Schneider, Th. Wolf, P. Dosanjh, R. Liang, and W.N. Hardy, Physica C {{\bf{317-318}}, 259 (1999).


\bibitem{Peters88:37} C. J. Peters, R. J. Birgeneau, M. A. Kastner, H. Yoshizawa, Y. Endoh, J. Tranquada, G. Shirane, Y. Hidaka, M. Oda, M. Suzuki, and T. Murakami, Phys. Rev. B {\bf{37}}, 9761 (1988).

\bibitem{Tranquada89:40} J. M. Tranquada, G. Shirane, B. Keimer, S. Shamoto, and M. Sato, Phys. Rev. B {\bf{40}}, 4503 (1989).

\bibitem{Lee99:69} C.-H. Lee, K. Yamada, Y. Endoh, G. Shirane, R.J. Birgeneau, M.A. Kastner, M. Greven, and Y.-J. Kim, J. Phys. Soc. Jpn. {\bf{69}}, 1170 (2000). 

\bibitem{Hiraka01:70} H. Hiraka, Y. Endoh, M. Fujita, Y.S. Lee, J. Kulda, A. Ivanov, and R.J. Birgeneau, J. Phys. Soc. Jpn. {\bf{70}}, 853 (2001).

\bibitem{Mason93:71} T. E. Mason, G. Aeppli, S. M. Hayden, A. P. Ramirez, and H. A. Mook, Phys. Rev. Lett. {\bf{71}}, 919 (1993).

\bibitem{Wells97:77} B.O. Wells, Y.S. Lee, M.A. Kastner, R.J. Christianson, R.J. Birgeneau, K. Yamada, Y. Endoh, and G. Shirane, Science {\bf{277}}, 1067 (1997).

\bibitem{Keimer92:46} B. Keimer, N. Belk, R. J. Birgeneau, A. Cassanho, C. Y. Chen, M. Greven, M. A. Kastner, A. Aharony, Y. Endoh, R. W. Erwin, and G. Shirane, Phys. Rev. B {\bf{46}}, 14034 (1992).

\bibitem{Keimer91:67} B. Keimer, R. J. Birgeneau, A. Cassanho, Y. Endoh, R. W. Erwin, M. A. Kastner, and G. Shirane, Phys. Rev. Lett. {\bf{67}}, 1930 (1991).

\bibitem{Aeppli97:278} G. Aeppli, T.E. Mason, S.M. Hayden, H.A. Mook, and J. Kulda, Science {\bf{278}}, 1432 (1997).

\bibitem{Bao03:91} W. Bao, Y. Chen, Y. Qiu, and J. L. Sarrao, Phys. Rev. Lett. {\bf{91}}, 127005 (2003).

\bibitem{Matusda93:5} M. Matsuda, R.J. Birgeneau, Y. Endoh, Y. Hidaka, M.A. Kastner, K. Nakajima, G. Shirane, T.R. Thurston, and K. Yamada, J. Phys. Soc. Jpn. {\bf{5}}, 1702 (1993).

\bibitem{Fong00:61} H. F. Fong, P. Bourges, Y. Sidis, L. P. Regnault, J. Bossy, A. Ivanov, D. L. Milius, I. A. Aksay, and B. Keimer, Phys. Rev. B {\bf{61}}, 14773-14786 (2000).

\bibitem{Dai01:63} P. Dai, H. A. Mook, R. D. Hunt, and F. Dogan, Phys. Rev. B {\bf{63}}, 054525 (2001).

\bibitem{Stock04:69} C. Stock, W. J. L. Buyers, R. Liang, D. Peets, Z. Tun, D. Bonn, W. N. Hardy, and R. J. Birgeneau, Phys. Rev. B {\bf{69}}, 014502 (2004).

\bibitem{Dai00:406} P. Dai, H.A. Mook, G. Aeppli, S.M. Hayden, and F. Dogan, Nature {\bf{406}}, 965 (2000).

\bibitem{Timusk99:62} T. Timusk and B. Statt. Rep. Prog. Phys. {\bf{62}}, 61 (1999).

\bibitem{Mignod92:180} J. Rossat-Mignod, L.P. Regnault, C. Vettier, P. Bourges, P. Burlet, J. Bossy, J.Y. Henry, and G. Lapertot, Physica B {\bf{180-181}}, 383 (1992).

\bibitem{Mook00:404} H.A. Mook, P. Dai, F. Dogan, and R.D. Hunt, Nature {\bf{404}}, 729 (2000).

\bibitem{Bourges00:288} P. Bourges, Y. Sidis, H.F. Fong, L.P. Regnault, J. Bossy, A. Ivanov, and B. Keimer, Science {\bf{288}}, 1234 (2000).

\bibitem{Hinkov04:430} V. Hinkov, S. Pailhes, P. Bourges, Y. Sidis, A. Ivanov, A. Kulakov, C.T. Lin, D.P. Chen, C. Berhnard, B. Keimer, Nature, {\bf{430}}, 650 (2004).

\bibitem{Balatsky99:82} A.V. Balatsky and P. Bourges, Phys. Rev. Lett. {\bf{82}}, 5337 (1999).

\bibitem{Lee99:60} Y.S. Lee, R.J. Birgeneau, M.A. Kastner, Y. Endoh, S. Wakimoto, K. Yamada, R. W. Erwin, S.-H. Lee, and G. Shirane, Phys. Rev. B {\bf{60}}, 3643 (1999).

\bibitem{Bascones05:xx} E. Bascones and T.M. Rice, unpublished (cond-mat/0511661).

\bibitem{Stock06:73} C. Stock, W.J.L. Buyers, Z. Yamani, C.L. Broholm, J.-H. Chung, Z. Tun, R. Liang, D. Bonn, W.N. Hardy, and R.J. Birgeneau, Phys. Rev. B {\bf{73}}, 100504(R) (2006). 

\bibitem{Liang02:383}  R. Liang, D.A. Bonn, W.N. Hardy, Janice C. Wynn, K.A. Moler, L. Lu, S. Larochelle, L. Zhou, M. Greven, L. Lurio, and S.G.J. Mochrie,  Physica C {\bf{383}}, 1 (2002).

\bibitem{Peets02:15} D. C. Peets, R. Liang, C. Stock, W. J. L. Buyers, Z. Tun, L. Taillefer, R. J. Birgeneau, D. A. Bonn, and W. N. Hardy Journal of Superconductivity, 15, 531 (2002).

\bibitem{Birgeneau92:87} R.J. Birgeneau, R.W. Erwin, P.M. Gehring, M.A. Kastner, B. Kiemer, M. Sato, S. Shamoto, G. Shirane, and J.M. Tranquada, Z. Phys. B {\bf{87}}, 15 (1992).


\bibitem{Aeppli89:62} G. Aeppli, S. M. Hayden, H. A. Mook, Z. Fisk, S.-W. Cheong, D. Rytz, J. P. Remeika, G. P. Espinosa, and A. S. Cooper Phys. Rev. Lett. {\bf{62}}, 2052-2055 (1989).

\bibitem{Reznik96:53} D. Reznik, P. Bourges, H. F. Fong, L. P. Regnault, J. Bossy, C. Vettier, D. L. Milius, I. A. Aksay, and B. Keimer Phys. Rev. B {\bf{53}}, 14741 (R) (1996).

\bibitem{Hayden96:54} S. M. Hayden, G. Aeppli, T. G. Perring, H. A. Mook, and F. Dogan Phys. Rev. B {\bf{54}}, 6905 (R) (1996).

\bibitem{Hayden96:76} S. M. Hayden, G. Aeppli, H. A. Mook, T. G. Perring, T. E. Mason, S.-W. Cheong, and Z. Fisk Phys. Rev. Lett. {\bf{76}}, 1344 (1996).

\bibitem{Coldea01:86}  R. Coldea, S. M. Hayden, G. Aeppli, T. G. Perring, C. D. Frost, T. E. Mason, S.-W. Cheong, and Z. Fisk Phys. Rev. Lett. {\bf{86}}, 5377 (2001).

\bibitem{Arai99:83} M. Arai, T. Nishijima, Y. Endoh, T. Egami, S. Tajima, K. Tomimoto, Y. Shiohara, M. Takahashi, A. Garrett, and S. M. Bennington, Phys. Rev. Lett. {\bf{83}}, 608 (1999).

\bibitem{Stock05:71} C. Stock, W. J. L. Buyers, R. A. Cowley, P. S. Clegg, R. Coldea, C. D. Frost, R. Liang, D. Peets, D. Bonn, W. N. Hardy, and R. J. Birgeneau, Phys. Rev. B {\bf{71}}, 024522 (2005).

\bibitem{Hayden04:429} S. M. Hayden, H. A. Mook, Pengcheng Dai, T. G. Perring, F. Dogan, Nature, {\bf{429}}, 531 (2004).

\bibitem{Tranquada04:429} J. M. Tranquada, H. Woo, T. G. Perring, H. Goka, G. D. Gu, G. Xu, M. Fujita, K. Yamada, Nature, {\bf{429}}, 534 (2004).

\bibitem{Christensen04:93} N. B. Christensen, D. F. McMorrow, H. M. Ronnow, B. Lake, S. M. Hayden, G. Aeppli, T. G. Perring, M. Mangkorntong, M. Nohara, and H. Takagi, Phys. Rev. Lett. {\bf{93}}, 147002 (2004).

\bibitem{Pailhes04:93} S. Pailhes, Y. Sidis, P. Bourges, V. Hinkov, A. Ivanov, C. Ulrich, L. P. Regnault, and B. Keimer, Phys. Rev. Lett. {\bf{93}}, 167001 (2004).

\bibitem{Kee02:88} H.-Y. Kee, S. A. Kivelson, and G. Aeppli, Phys. Rev. Lett. {\bf{88}}, 257002 (2002).


\bibitem{Yamada98:57} K. Yamada, C. H. Lee, K. Kurahashi, J. Wada, S. Wakimoto, S. Ueki, H. Kimura, Y. Endoh, S. Hosoya, G. Shirane, R.J. Birgeneau, M. Greven, M.A. Kastner, and Y.J. Kim, Phys. Rev. B {\bf{57}}, 6165 (1998). 

\bibitem{Suzuki99:59} T. Suzuki, T. Goto, K. Chiba, T. Shinoda, T. Fukase, H. Kimura, K. Yamada, M. Ohashi, and Y. Yamaguchi, Phys. Rev. B 57, 3229 (1998); H. Kimura, K. Hirota, H. Matsushita, K. Yamada, Y. Endoh, S. H. Lee, C. F. Majkrzak, R. Erwin, G. Shirane, M. Greve, Y. S. Lee, M. A. Kastner, and R. J. Birgeneau, Phys. Rev. B {\bf{59}}, 6517 (1999).

\bibitem{Wakimoto99:60} S. Wakimoto, G. Shirane, Y. Endoh, K. Hirota, S. Ueki, K. Yamada, R. J. Birgeneau, M. A. Kastner, Y. S. Lee, P. M. Gehring, and S. H. Lee, Phys. Rev. B {\bf{60}}, 769 (R) (1999).

\bibitem{Wakimoto00:61} S. Wakimoto, R. J. Birgeneau, M. A. Kastner, Y. S. Lee, R. Erwin, P. M. Gehring, S. H. Lee, M. Fujita, K. Yamada,Y. Endoh, K. Hirota, and G. Shirane, Phys. Rev. B {\bf{61}}, 3699 (2000).

\bibitem{Matsuda00:62} M. Matsuda, M. Fujita, K. Yamada, R. J. Birgeneau, M. A. Kastner, Y. Endoh, S. Wakimoto, and G. Shirane, Phys. Rev. B {\bf{62}}, 9148 (2000).

\bibitem{Matsuda02:65} M. Matsuda, M. Fujita, K. Yamada, R. J. Birgeneau, Y. Endoh, and G. Shirane, Phys. Rev. B {\bf{65}}, 134515 (2002).

\bibitem{Endoh88:37} Y. Endoh, K. Yamada, R. J. Birgeneau, D. R. Gabbe, H. P. Jenssen, M. A. Kastner, C. J. Peters, P. J. Picone, T. R. Thurston, J. M. Tranquada, G. Shirane, Y. Hidaka, M. Oda, Y. Enomoto, M. Suzuki and T. Murakami Phys. Rev. B {\bf{37}}, 7443 (1988).

\bibitem{Fujita02:65} M. Fujita, K. Yamada, H. Hiraka, P. M. Gehring, S. H. Lee, S. Wakimoto, and G. Shirane, Phys. Rev. B {\bf{65}}, 064505 (2002).

\bibitem{Nied98:80} Ch. Niedermayer, C. Bernhard, T. Blasius,  A. Golnik, A. Moodenbaugh, and J. I. Budnick, Phys. Rev. Lett. {\bf{80}}, 3843 (1998).

\bibitem{Ichikawa00:85} N. Ichikawa,  S. Uchida, J. M. Tranquada, T. Niemoeller, P. M. Gehring, S.-H. Lee, J. R. Schneider, Phys. Rev. Lett. {\bf{85}}, 1738 (2000).

\bibitem{Fujita02:88} M. Fujita, H. Goka, K. Yamada, and M. Matsuda, Phys. Rev. Lett. {\bf{88}}, 167008 (2002). 

\bibitem{Hasselmann01:56} N. Hasselmann, A. H. Castro Neto, and C. M. Smith, Europhys. Lett. {\bf{56}}, 870 (2001).

\bibitem{Hasselmann04:69} N. Hasselmann, A. H. Castro Neto, and C. M. Smith, Phys. Rev. B {\bf{69}}, 014424 (2004).

\bibitem{Matsuda06:xx} M. Matsuda, M. Fujita, K. Yamada, Phys. Rev. B {\bf{73}}, 14503 (R) (2006).

\bibitem{Hiraka05:74} H. Hiraka, T. Machi, N. Watanabe, Y. Itoh, M. Matsuda, and K. Yamada, J. Phys. Soc. Jpn. {\bf{74}}, 2197 (2005).


\bibitem{Lee04:69} Y.S. Lee, F.C. Chou, A. Tewary, M.A. Kastner, S.H. Lee, and R.J. Birgeneau, Phys. Rev. B {\bf{69}}, 020502 (2004).

\bibitem{Katano00:62} S. Katano, M. Sato, K. Yamada, T. Suzuki, and T. Fukase, Phys. Rev. B {\bf{62}}, R14677 (2000).

\bibitem{Lake01:291} B. Lake, G. Aeppli, K. N. Clausen, D. F. McMorrow, K. Lefmann, N. E. Hussey, N. Mangkorntong, M. Nohara, H. Takagi, T. E. Mason, and A. Schroader, Science {\bf{291}}, 1759 (2001).

\bibitem{Lake02:299} B. Lake, H. M. Ronnow, N. B. Christensen, G. Aeppli, K. Lefmann, D. F. McMorrow, P. Vorderwisch, P. Smeibidl, N. Mangkorntong, T. Sasagawa, M. Nohara, H. Takagi, and T. E. Mason, Nature {\bf{415}}, 299 (2002).

\bibitem{Khaykovich02:66} B. Khaykovich, Y. S. Lee, R. W. Erwin, S.-H. Lee, S. Wakimoto, K. J. Thomas, M. A. Kastner, and R. J. Birgeneau, Phys. Rev. B {\bf{66}}, 014528 (2002). 

\bibitem{Matsuda02:66} M. Matsuda, M. Fujita, K. Yamada, R.J. Birgeneau, Y. Endoh, and G. Shirane, Phys. Rev. B {\bf{66}}, 174508 (2002).

\bibitem{Wakimoto03:67} S. Wakimoto,  R. J. Birgeneau, Y. Fujimaki, N. Ichikawa, T. Kasuga, Y. J. Kim, K. M. Kojima, S.-H. Lee, H. Niko, J. M. Tranquada, S. Uchida, and M. v. Zimmermann, Phys. Rev. B {\bf{67}}, 184419 (2003).

\bibitem{Khaykovich05:71} B. Khaykovich, S. Wakimoto, R. J. Birgeneau, M. A. Kastner, Y. S. Lee, P. Smeibidl, P. Vorderwisch, and K. Yamada, Phys. Rev. B {\bf{71}}, 220508 (2005).

\bibitem{Lake05:4} B. Lake, K. Lefmann, N.B. Christensen, G. Aeppli, D.F. McMorrow, N.M. Ronnow, P. Vorderwisch, P. Smeibidl, N. Mangkorntong, T. Sasagawa, M. Nohara, and H. Takagi, Nat. Mat. {\bf{4}}, 658 (2005).

\bibitem{Demler01:87} E. Demler, S. Sachdev, and Y. Zhang, Phys. Rev, Lett. {\bf{87}}, 67202, (2001).

\bibitem{Zhang02:66} Y. Zhang, E. Demler, and S. Sachdev, Phys. Rev. B {\bf{66}}, 94501 (2002).

\bibitem{Arovas97:79} D.P. Arovas, A. J. Berlinsky, C. Kallin, and Shou-Cheng Zhang, Phys. Rev. Lett. {\bf{79}}, 2871 (1997).

\bibitem{Kivelson01:98} S. A. Kivelson, G. Aeppli, and V.J. Emery, PNAS {\bf{98}}, 11903 (2001).


\bibitem{zaan01}
J. Zaanen, O.~Y. Osman, H.~V. Kruis, Z. Nussinov, and J. {Tworzyd\l o}, Phil.
  Mag. B {\bf 81},  1485  (2001).

\bibitem{sach91}
S. Sachdev and N. Read, Int. J. Mod. Phys. B {\bf 5},  219  (1991).

\bibitem{mach89}
K. Machida, Physica C {\bf 158},  192  (1989).

\bibitem{cast95}
C. Castellani, C. {Di Castro}, and M. Grilli, Phys. Rev. Lett. {\bf 75},  4650
  (1995).

\bibitem{tran95a}
J.~M. Tranquada, B.~J. Sternlieb, J.~D. Axe, Y. Nakamura, and S. Uchida, Nature
  {\bf 375},  561  (1995).

\bibitem{fuji04}
M. Fujita, H. Goka, K. Yamada, J.~M. Tranquada, and L.~P. Regnault, Phys. Rev.
  B {\bf 70},  104517  (2004).

\bibitem{abba05}
P. Abbamonte, A. Rusydi, S. Smadici, G.~D. Gu, G.~A. Sawatzky, and D.~L. Feng,
  Nature Physics {\bf 1},  155  (2005).

\bibitem{ulri02}
C. Ulrich, S. Kondo, M. Reehuis, H. He, C. Bernhard, C. Niedarmayer, F.
  Bour\'ee, P. Bourges, M. Ohl, H.~M. R{\o}nnow, H. Takagi, and B. Keimer,
  Phys. Rev. B {\bf 65},  220507  (2002).

\bibitem{huck05}
M. H\"ucker, Y.-J. Kim, G.~D. Gu, J.~M. Tranquada, B.~D. Gaulin, and J.~W.
  Lynn, Phys. Rev. B {\bf 71},  094510  (2005).

\bibitem{wang03}
N.~L. Wang, P. Zheng, T. Feng, G.~D. Gu, C.~C. Homes, J.~M. Tranquada, B.~D.
  Gaulin, and T. Timusk, Phys. Rev. B {\bf 67},  134526  (2003).
\bibitem{waki04}
S. Wakimoto, H. Zhang, K. Yamada, I. Swainson, H. Kim, and R.~J. Birgeneau,
  Phys. Rev. Lett. {\bf 92},  217004  (2004).

\bibitem{waki06}
S. Wakimoto, R.~J. Birgeneau, C.~D. Frost, A. Kagedan, H.~K. Kim, I. Swainson,
  J.~M. Tranquada, K. Yamada, and H. Zhang, unpublished.

\bibitem{bern01b}
C. Bernhard, J.~L. Tallon, T. Blasius, A. Golnik, and C. Niedermayer, Phys.
  Rev. Lett. {\bf 86},  1614  (2001).

\bibitem{uemu01}
Y.~J. Uemura, Solid State Commun. {\bf 120},  347  (2001).

\bibitem{wen00}
H.~H. Wen, X.~H. Chen, W.~L. Yang, and Z.~X. Zhao, Phys. Rev. Lett. {\bf 85},
  2805  (2000).
\bibitem{rezn06}
D. Reznik, L. Pintschovius, M. Ito, S. Iikubo, M. Sato, H. Goka, M. Fujita, K.
  Yamada, G.~D. Gu, and J.~M. Tranquada, Nature {\bf 440},  1170  (2006).

\bibitem{bohn03}
K.-P. Bohnen, R. Heid, and M. Krauss, Europhys. Lett. {\bf 64},  104  (2003).

\bibitem{kive05a}
S.~A. Kivelson and E. Fradkin, cond-mat/0507459.

\bibitem{kao00}
Y.-J. Kao, Q. Si, and K. Levin, Phys. Rev. B {\bf 61},  R11898  (2000).

\bibitem{norm00}
M.~R. Norman, Phys. Rev. B {\bf 61},  14751  (2000).

\bibitem{brin99}
J. Brinckmann and P.~A. Lee, Phys. Rev. Lett. {\bf 82},  2915  (1999).

\bibitem{chub01}
A.~V. Chubukov, B. Jank\'o, and O. Tchernyshyov, Phys. Rev. B {\bf 63},
  180507(R)  (2001).

\bibitem{onuf02}
F. Onufrieva and P. Pfeuty, Phys. Rev. B {\bf 65},  054515  (2002).
\end{document}